\documentclass[]{spie}  

\usepackage{caption}
\usepackage{subfig}
\usepackage{amsmath,amsfonts,amssymb}
\usepackage{graphicx} 
\usepackage{color} 
\usepackage[colorlinks=true, allcolors=blue]{hyperref} 
\usepackage{listings} 
\usepackage{textcomp} 
\usepackage{verbatim} 
\newcommand{\jovo}[1]{{\color{red}{\it }}}
\title{Deformably Registering and Annotating Whole CLARITY Brains to an Atlas via Masked LDDMM}

\author[a]{Kwame S. Kutten}
\author[a,c]{Joshua T. Vogelstein}
\author[b]{Nicolas Charon}
\author[d]{Li Ye}
\author[d]{Karl Deisseroth}
\author[a]{Michael I. Miller}
\affil[a]{Department of Biomedical Engineering, Johns Hopkins University}
\affil[b]{Department of Applied Mathematics \& Statistics, Johns Hopkins University}
\affil[c]{Institute for Computational Medicine, Johns Hopkins University}
\affil[d]{Department of Bioengineering, Stanford University}

\authorinfo{Further author information: (Send correspondence to K.S.K.)\\K.S.K.: E-mail: kkutten1@jhmi.edu}

\graphicspath{{./images/}}
\begin{document} 
\noindent Kwame S. Kutten, Joshua T. Vogelstein, Nicolas Charon, Li Ye, Karl Deisseroth, Michael I. Miller,
``Deformably registering and annotating whole CLARITY brains to an atlas via masked LDDMM,''
Optics, Photonics and Digital Technologies for Imaging Applications IV, 
Peter Schelkens, Touradj Ebrahimi, Gabriel Crist\'{o}bal, Fr\'{e}d\'{e}ric Truchetet, Pasi Saarikko, Editors,
Proc. SPIE 9896, 989616 (2016).

\noindent Copyright 2016 Society of Photo Optical Instrumentation Engineers.
One print or electronic copy may be made for personal use only.
Systematic electronic or print reproduction and distribution, duplication of any material in this paper for a fee or for commercial purposes, or modification of the content of the paper are prohibited.

\noindent \url{http://dx.doi.org/10.1117/12.2227444}

\newpage

\maketitle

\begin{abstract}
The CLARITY method renders brains optically transparent to enable high-resolution imaging in the structurally intact brain.
Anatomically annotating CLARITY brains is necessary for discovering which regions contain signals of interest.
Manually annotating whole-brain, terabyte CLARITY images is difficult, time-consuming, subjective, and error-prone.
Automatically registering CLARITY images to a pre-annotated brain atlas offers a solution, but is difficult for several reasons.
Removal of the brain from the skull and subsequent storage and processing cause variable non-rigid deformations, thus compounding inter-subject anatomical variability.
Additionally, the signal in CLARITY images arises from various biochemical contrast agents which only sparsely label brain structures.
This sparse labeling challenges the most commonly used registration algorithms that need to match image histogram statistics to the more densely labeled histological brain atlases.
The standard method is a multiscale Mutual Information B-spline algorithm that dynamically generates an average template as an intermediate registration target. 
We determined that this method performs poorly when registering CLARITY brains to the Allen Institute's Mouse Reference Atlas (ARA), because the image histogram statistics are poorly matched.
Therefore, we developed a method (Mask-LDDMM) for registering CLARITY images, that automatically finds the brain boundary and learns the optimal deformation between the brain and atlas masks.
Using Mask-LDDMM without an average template provided better results than the standard approach when registering CLARITY brains to the ARA.
The LDDMM pipelines developed here provide a fast automated way to anatomically annotate CLARITY images; our code is available as open source software at \url{http://NeuroData.io}.

\end{abstract}


\section{INTRODUCTION}
\label{sec:intro}  

Imaging whole brains at the cellular level without disturbing their underlying structure has always been challenging.
All cells are surrounded by a phospholipid bilayer which scatters light, rendering most biological tissues opaque to the naked eye.
Thus it is often necessary to physically slice brains in order to use light microscopy.
Sectioning tissue has two major drawbacks for researchers interested in building a whole brain connectome.
First, slicing can dislocate synapses and axons necessary for tracing neuronal circuitry.
Second, the inter-sectional resolution will always be much lower than the intra-sectional resolution, making neurite tracing difficult \cite{Kim}.

\subsection{CLARITY}
\begin{figure}[t]
 \begin{center}
  \begin{tabular}{c c}
   \includegraphics[height=2.5in]{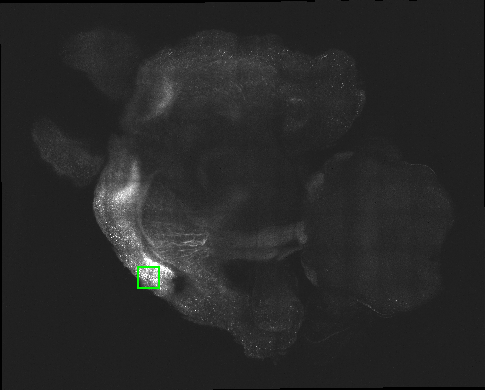} & \includegraphics[height=2.5in]{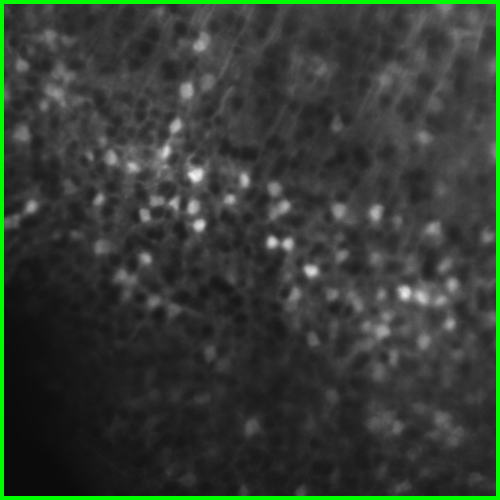} 
  \end{tabular}
 \end{center}
 \caption{Axial slice through CLARITY image and magnified cutout acquired by light-sheet microscopy.}
 \label{fig:clarity}
\end{figure}

CLARITY avoids these problems by converting the brain into a translucent hydrogel-tissue hybrid.
In the procedure, the brain is first perfused with hydrogel monomers and formaldehyde.
When heated, the monomers and formaldehyde polymerize to form a molecular mesh which crosslinks amine groups of biological molecules.
Since phospholipids lack amine groups, they do not crosslink with the mesh and can be eluted away with a strong detergent.
The remaining hydrogel-brain hybrid is relatively translucent and permeable to fluorescent antibodies, making it amenable to labeling and interrogation by light-sheet microscopy. \cite{Chung, Tomer}
An axial slice through a CLARITY volume and magnified cutout are shown in Fig.~\ref{fig:clarity}.

\subsection{NeuroData Cluster}

CLARITY image volumes are often over 1 terabyte in size, far too large to be visualized and analyzed on a personal computer.
Furthermore, extracting meaningful information from ``big'' images can be both time-consuming and difficult.
Thus, the NeuroData cluster was created to address these challenges.
As part of the project, open source software for access, visualization, and analysis of terabyte-scale images was developed.
Images are stored in a multiresolution hierarchy with level 0 being the native resolution, and each subsequent level at half the previous level's resolution.
Thus the infrastructure is optimized for applying computer vision algorithms in parallel across multiple scales.\cite{Burns13}

Data can be accessed through a RESTful API, a stateless interface which allows end users to download image cutouts or upload data using specific URLs.
%
%
%
In the API, images are identified by a unique token, each of which can have one or more channels.\cite{Burns13}
The \emph{Connectome Annotation for Joint Analysis of Large data (CAJAL)} package provides access to this API through MATLAB \cite{GrayRoncal} while \emph{NeuroData Input/Output (ndio)} provides access through Python.
Images can also be visualized in a web browser using NeuroDataViz.

\subsection{Motivation}
\begin{figure}[t]
 \begin{center}
  \includegraphics[width=\textwidth]{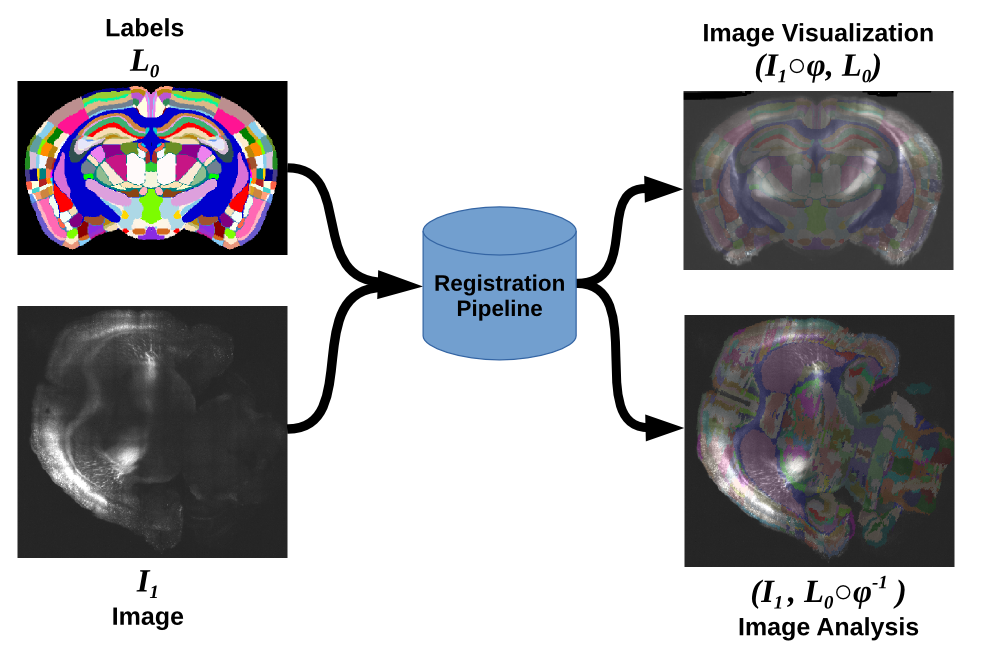} \\
 \end{center}
 \caption{The registration pipeline transforms the atlas labels to the CLARITY image's space for image analysis and the CLARITY image to an atlas space for visualization.}
 \label{fig:workflow}
\end{figure}

By annotating an entire image volume, one can draw conclusions on the texture and shape of a given brain structure.
Since manual labeling is time-consuming, the most efficient annotation method is registration to a standard atlas.
Spatially normalizing several subject brains into an atlas space makes it easier to determine how one brain differs from another in any given structure or location.
Furthermore, this can also aide in visualization.
Raw CLARITY images are often acquired in an oblique plane, making it difficult for observers to identify structures on a 2D display.
Aligning the brains with an atlas solves this problem by allowing brain visualization in one of the three standard planes (axial, sagittal, and coronal).
Transforming atlas labels to a subject's space facilitates analysis of image features within brain structures. (Fig.~\ref{fig:workflow}).

\subsection{Previous Work}
Registering CLARITY images to each other or to an atlas has recently become a topic of interest.
A preceding study described the development of a pipeline for registering the Allen Institute's Mouse Reference Atlas (ARA) to images of transsynaptic viral traced brains. \cite{Kutten}
The ARA is a widely used mouse brain atlas which includes Nissl-stained reference images and over 700 manually defined brain structures.\cite{ABAWhitePaper}
A test image was acquired using serial-two photon (STP) tomography, a technique which pairs a two-photon microscope with a vibratome for automatic tissue slicing. \cite{Ragan}
Since the intensity profiles of the Nissl-stained ARA and the test image differed greatly, the images were registered using their corresponding brain masks.
Masks were aligned first using affine registration, followed by deformable registration using Large Deformation Diffeomorphic Metric Mapping (LDDMM), an algorithm which computes smooth invertible transforms between image volumes. 
Qualitative observation showed that the registration results were acceptable in most parts of the brain, although alignment of deeper structures were less accurate.\cite{Kutten}
In a different study, 25 CLARITY images were registered and averaged to create a single reference template using a mutual information metric with B-spline transforms.  This template was then used to construct an atlas for experiments combining CLARITY with transsynaptic viral tracing. \cite{Menegas}

\subsection{ITK}
The \emph{Insight Segmentation and Registration Toolkit} (ITK) is an open source library funded by the National Library of Medicine.
It was developed by Kitware Inc. and has been used widely within the medical imaging community. \cite{ITK1}
ITK's registration framework is designed to be modular.  To register a moving image to a fixed image, the user selects a metric (e.g. mean squared error) to compare the images, and the transform (e.g. affine) applied to the moving image. An optimizer (e.g. gradient descent) is used iteratively to improve the transform parameters.
%
%
Additionally, the user can select the type of interpolation (e.g. nearest neighbor) for resampling the moving image. \cite{ITK2}
ITK version 4 includes support for time-varying velocity field transforms often used in diffeomorphic registration algorithms such as LDDMM.
This greatly facilitates the implementation of these types of algorithms in ITK. \cite{Avants12}
This functionality can be used in Python through SimpleITK, an easy-to-use interface for ITK's algorithms. \cite{Lowekamp}



\subsection{Challenges}
\begin{figure}[ht]
 \centering
 \subfloat[]{\label{fig:cerebralPeduncle} \includegraphics[height=1.20in]{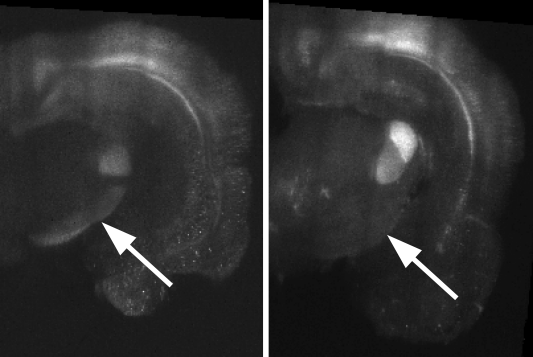} }
 \subfloat[]{\label{fig:deformed}\includegraphics[height=1.20in]{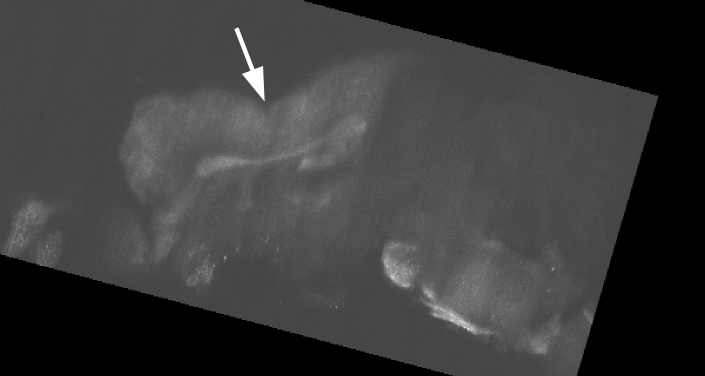} }
 \subfloat[]{\label{fig:missingData}\includegraphics[height=1.20in]{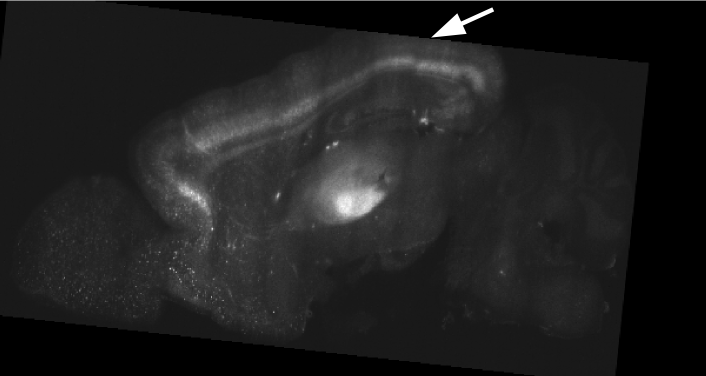} }
 \caption{Arrows point out features which make registering CLARITY images challenging.
          In (a) the cerebral peduncle is light in one CLARITY brain but dark in another.
          In (b) the brain was greatly deformed in the clarifying process and in (c) the brain is missing data.}
 \label{fig:challenges}
\end{figure}

In addition to their large size, CLARITY images present several unique challenges for image registration.
In the CLARITY images of this study, a neuron's brightness is proportional to its activity, which means that CLARITY images have a functional component.
Regions which appear bright in one CLARITY brain may appear dark in another (Fig.~\ref{fig:cerebralPeduncle}).
Registration is further complicated by brain deformation introduced in the clarifying process (Fig.~\ref{fig:deformed}) and missing data (Fig.~\ref{fig:missingData}).

\section{METHODS}
The registration pipeline from the preceding study \cite{Kutten} was reimplemented using SimpleITK.
This pipeline, known henceforth as the Mask-LDDMM pipeline, registered images using their masks.
Additionally, an Image-LDDMM pipeline which directly registered images using intensity was developed.


\subsection{Image Acquisition}
12 CLARITY mouse brains (5 wild type controls and 7 behaviorally challenged) were imaged using CLARITY-Optimized Light-sheet Microscopy (COLM) (whole brain COLM imaging and data stitching performed by R. Tomer, in preparation).
In brief, raw volumes were acquired in 0.585 $\mu$m x 0.585 $\mu$m resolution slices with a slice spacing of 5 to 8 $\mu$m.
Images were stored in the NeuroData cluster at 6 resolution levels, with level 0 being the full resolution and level 5 being the lowest resolution.
To avoid registration complications, four CLARITY brains which were not missing any data (Control239, Challenged178, Challenged199, and Challenged188) were selected to test the pipelines.

\subsection{Mask Generation}
Let $\Omega \subset \mathbb{R}^3$ be the background space.
Let $I_0: \Omega \rightarrow \mathbb{R}$ be the template (moving) image which will be deformed to match target (fixed) image $I_1: \Omega \rightarrow \mathbb{R}$.
Thus for CLARITY-ARA registrations $I_0$ is the the Nissl-stained ARA, $L_0: \Omega \rightarrow \mathbb{N}$ are the corresponding ARA annotations, and $I_1$ is a CLARITY image.
Since the resolution of ARA version 2 is 25 $\mu$m isotropic, CLARITY images were downloaded from the NeuroData cluster and resampled to the same resolution.
ARA mask $M_0: \Omega \rightarrow \{0,1\}$ was generated from $L_0$ by taking the union of all foreground labels.
For each CLARITY image $I_1$, mask $M_1: \Omega \rightarrow \{0,1\}$ was generated using the following procedure.
First, $I_1$ was binary-thresholded to remove the background.
Next, this rough mask was opened using a 50 $\mu$m radius ball-shaped kernel to remove foreground grains.
Finally, the mask was closed by a 125 $\mu$m radius kernel to remove large foreground holes.

\subsection{Preprocessing}
In the \emph{Image-LDDMM Pipeline}, preprocessing consisted of two steps. 
In the first step, the brain masks $M_0$ and $M_1$ were applied to the images $I_0$ and $I_1$ respectively.
Next, the masked $I_0$ was histogram-matched to the masked $I_1$.
In the matching procedure, 32-bin histograms were calculated for both the template and target images.
The histograms were matched exactly at 8 quantile points, and by interpolation at all other intensities between these points.
In the \emph{Mask-LDDMM pipeline}, no preprocessing was done.
Instead $I_0$ and $I_1$ were replaced by their corresponding masks, $M_0$ and $M_1$, during registration.

\subsection{Registration}
Registration was then done in a three step process with rigid alignment, followed by affine alignment, and finally deformable registration.
In rigid registration, parameters for the rigid transformation augmented matrix $R \in \mathbb{R}^{4 \times 4}$ were optimized using gradient descent on the mean squared error between the transformed histogram-matched $I_0$ and $I_1$.
For affine alignment, the same optimization scheme and image metric were used to find affine matrix $A \in \mathbb{R}^{4 \times 4}$ between $I_0 \circ R^{-1}$ and $I_1$.
The results of these intermediate steps, the Mask-Rigid/Mask-Affine in the Mask-LDDMM pipeline and Image-Rigid/Image-Affine in the Image-LDDMM pipeline, were stored for quantitative evaluation.

Let $J_0 = I_0 \circ R^{-1}A^{-1}$ and $J_1 = I_1$. Deformable registration was done by the LDDMM algorithm which used gradient descent to minimize the objective function
\begin{equation*}
 E(v) = \int_0^1 || L v(t) ||^2_{L_2} dt + \frac{1}{\sigma^2} ||J_0 \circ \phi_{10} - J_1||^2_{L_2}
\end{equation*}
where $v : [0,1] \times \Omega \rightarrow \mathbb{R}^3$ is the velocity of the flow from $I_0$ to $I_1$ and $L = (-\alpha \Delta + \gamma) \mathbf{I_3}$ is a kernel which ensures that $v$ is sufficiently smooth.
The greater $\alpha \in (0, \infty)$ is, the smoother the transform. \cite{Beg05}.
By convention $\gamma=1$ and $\sigma=1$ were used.
The algorithm was implemented in ITK by building upon the library's time-varying velocity field registration method.
To minimize computational cost, velocity field $v$ was discretized into only 4 time steps.
Displacement $\phi_{10} = \int_0^1 v(t,\phi_{10}) dt $ was found by integrating $v$ using a 4th order Runge-Kutta algorithm.
After LDDMM the final transform $\varphi = \phi_{01}AR$ and its inverse $\varphi^{-1} = R^{-1}A^{-1}\phi_{10}$  were found.
\subsection{Postprocessing}
\begin{figure}[t]
 \begin{center}
  \begin{tabular}{c c}
   \includegraphics[width=0.4\textwidth]{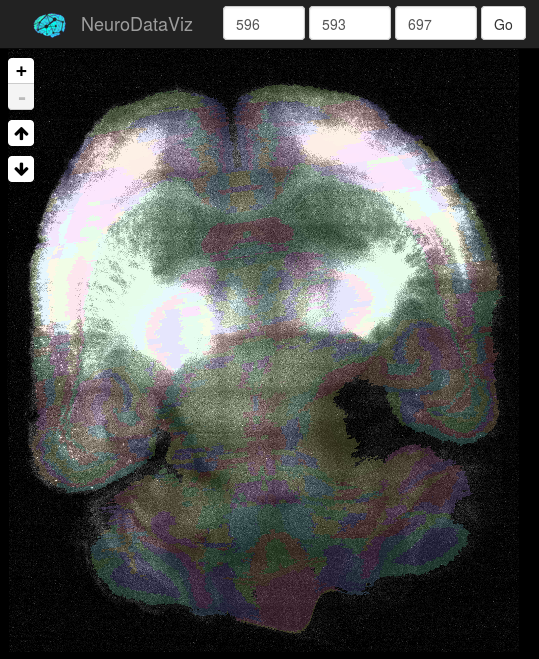} & \includegraphics[width=0.4\textwidth]{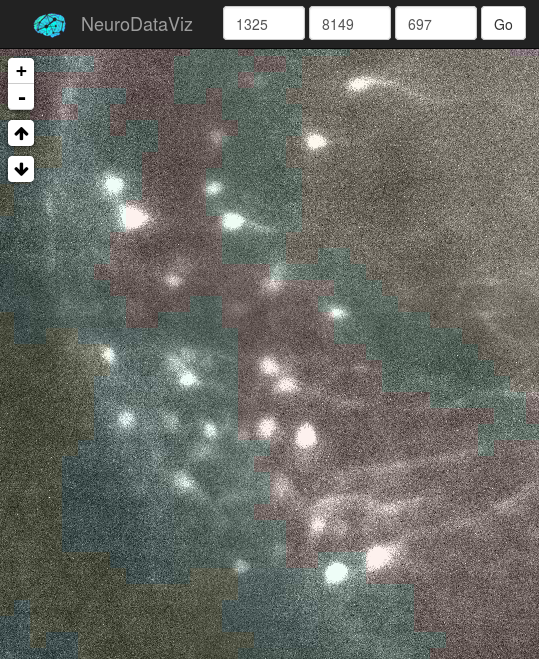} 
  \end{tabular}
 \end{center}
 \caption{Viewing CLARITY brain Challenged199 with overlay of ARA labels from the Mask-LDDMM pipeline in a web browser using NeuroDataViz.}
 \label{fig:ndviz}
\end{figure}

Deformed labels $L_0 \circ \varphi^{-1}$ were resampled to a level 5 resolution and fed into the NeuroData cluster.
The infrastructure automatically propagated the annotations to higher resolution levels.
Thus the full resolution images with ARA labels overlaid could be visualized from a web browser in NeuroDataViz (Fig.~\ref{fig:ndviz}).

\subsection{NeuroData Registration in Python}
Image processing and registration code from the pipelines were turned into a python module called \emph{NeuroData Registration (ndreg)}.
Functions from ndreg store results as NIfTI image files to leave a record of each step for easy debugging.
Internally, these functions called SimpleITK or custom binaries.
The module also includes convenience functions for downloading and uploading data to the NeuroData cluster through ndio.
The ndreg module is available as open source software at \url{http://NeuroData.io}.







\subsection{Quantitative Evaluation}
The registrations were quantitatively evaluated using aligned template-to-target mutual information, surface error, and landmark error.
Let $T$ and $U$ be random variables representing the intensities of deformed template image $I_0 \circ \varphi^{-1}$ and target image $I_1$ respectively.
The \emph{Mutual Information (MI)} between these images is
\begin{equation*}
 MI(T,U) = \iint_{\mathbb{R}\times\mathbb{R}} p(t,u) \ln \left(\frac{p(t,u)}{p(t)p(u)}\right)dt du
\end{equation*}
where $p(t)$, $p(u)$, and $p(t,u)$ are the deformed template histogram, target histogram, and joint histogram respectively.
Computing MI directly may yield unstable results.
Therefore it was estimated using the \emph{Viola-Wells} method, as implemented in ITK.
As recomended in the doccumentation, a standard deviation of 0.4 was used to smooth the histograms after normalizing the template and target image intensities to a zero mean and unity standard deviation.
Densities $p(t)$, $p(u)$, and $p(t,u)$  were then estimated from 1000 foreground samples using a Gaussian distribution-based Parzen window.


Landmark-based methods were also used to evaluate the results.  
Specifically, $N = 55$ landmarks were chosen and placed on the ARA and CLARITY images using the MRI Studio software suite's DiffeoMap program (\url{https://www.mristudio.org/}).
Of the landmarks, 28 were placed on the surface of the brain while the remaining 27 were placed on internal structures.
The \emph{error in position of the $k^{th}$ landmark} after registration is
\begin{equation*}
 e_k = d(\varphi(x_k), y_k) = || \varphi(x_k) - y_k ||_{L_2}
\end{equation*}
where $x_k$ and $y_k$ are the positions of the $k^{th}$ template and target landmarks for $k \in \{1,..., N\}$.

The \emph{Hausdorff Distance (HD)} between the deformed template and target surfaces is defined as
\begin{equation*}
 d_H(S_1,S(1)) = max \left\{ \sup_{x \in S(1)} \inf_{y \in S_1} d(x,y), \sup_{y \in S_1} \inf_{x \in S(1)} d(x,y) \right\}
\end{equation*}
where $S(1) = \partial(M_0 \circ \varphi^{-1})$ and $S_1 = \partial M_1$ denote the surfaces of the deformed template and target respectively.
Since the Hausdorff Distance is a maximum distance between surfaces, it was usually too sensitive to outlier surface points.
%
%
%
Therefore it was more convenient to compute the \emph{median of the surface distances between all points in $S_1$ and $S(1)$}
\begin{equation*}
 d_M(S_1, S(1)) = median\left\{\inf_{y \in S_1} d(x,y) \} \cup  \{\inf_{x \in S(1)} d(x,y)\right\}.
\end{equation*}


\section{RESULTS}

\begin{figure}[ht]
 \centering
 \subfloat[]{\includegraphics[width=0.33\textwidth]{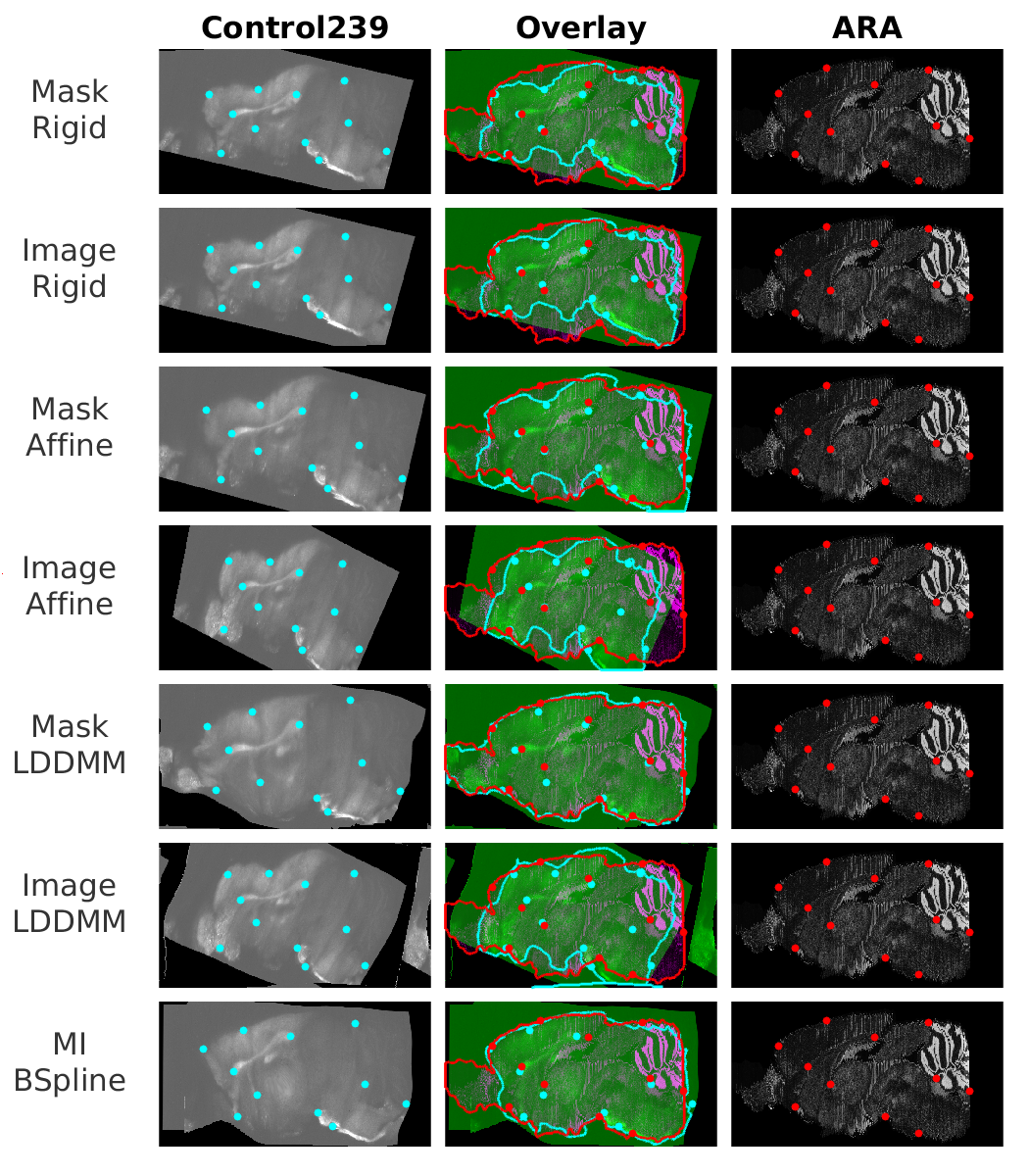} \label{fig:clarityToAra}}
 \subfloat[]{\includegraphics[width=0.33\textwidth]{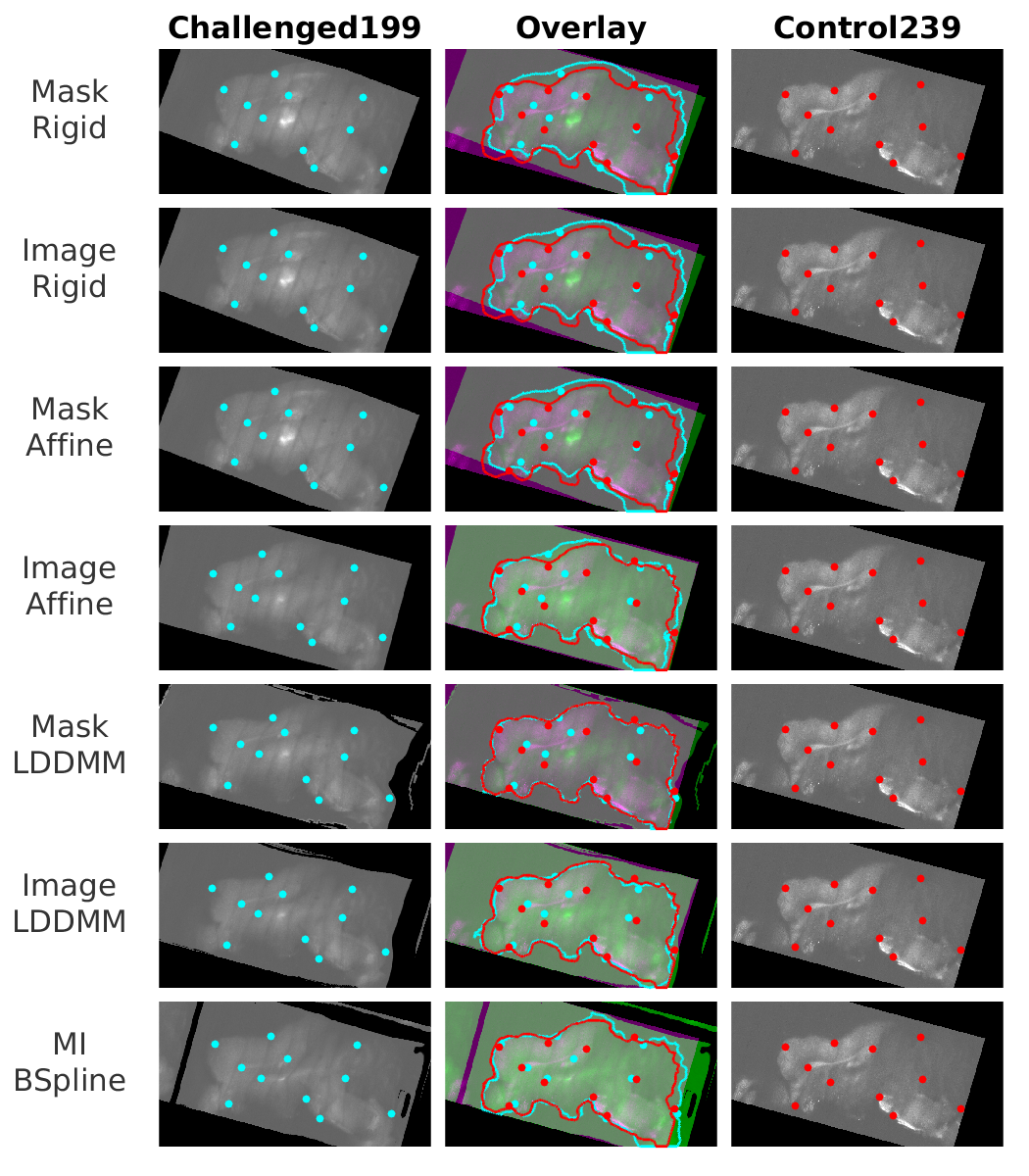} \label{fig:betweenConditions}} 
 \subfloat[]{\includegraphics[width=0.33\textwidth]{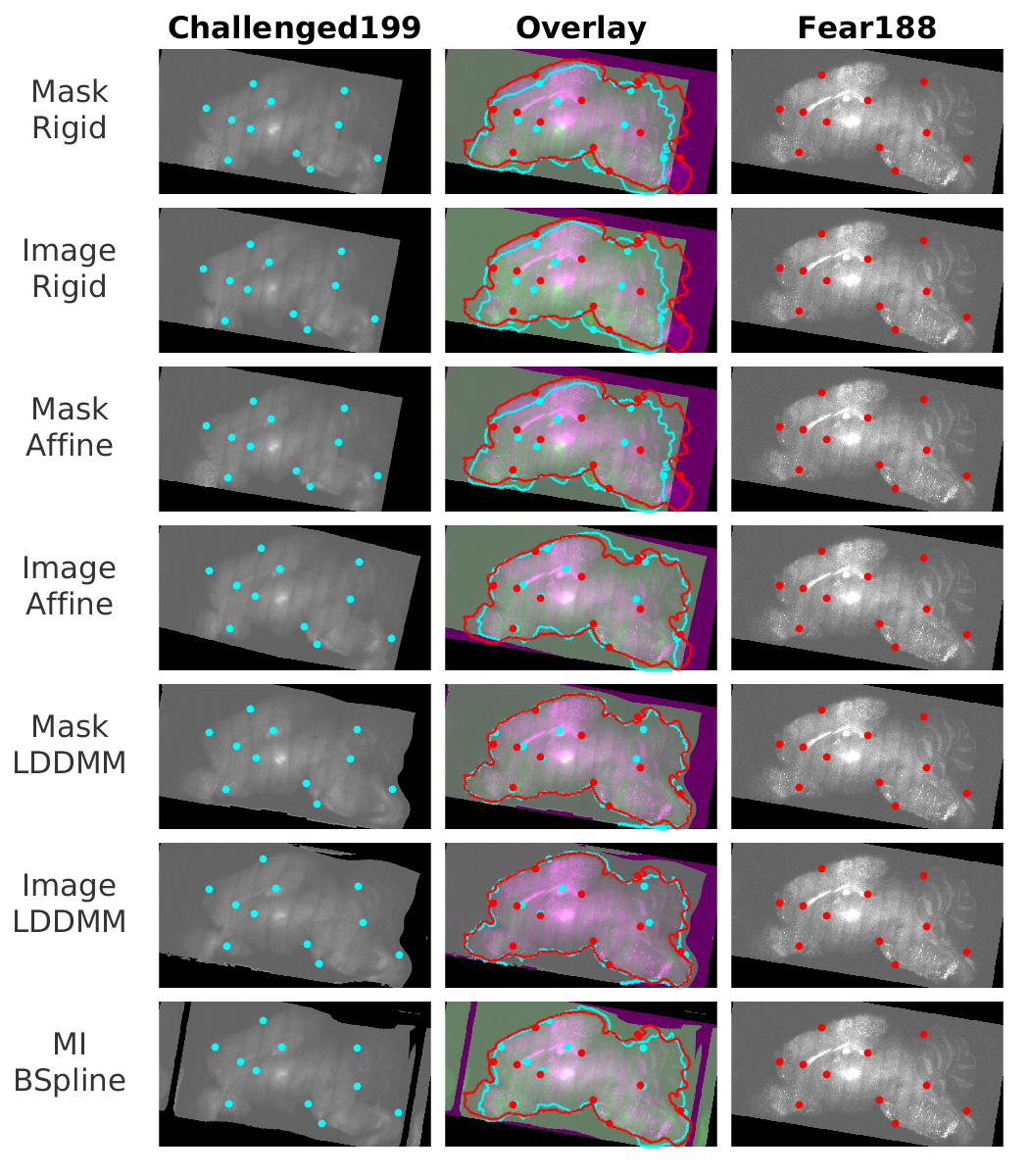} \label{fig:withinConditions}}
 \vspace{1cm}
 \caption{ Selected alignments from the CLARITY-ARA experiment (a), the CLARITY-CLARITY experiment between Challenged and Control (b), and the CLARITY-CLARITY experiment within Challenged (c).
   In each, the left column shows the template image and landmarks (in cyan). The right column shows the target image and landmarks (in red).
   The middle column shows the aligned template image, surface, and landmarks overlaid on the target image, surface, and landmarks. }
 \label{fig:expriments}
\end{figure}


Three experiments were performed. 
In the first experiment, CLARITY images were registered directly to the ARA.
In the next experiment, CLARITY images were registered to other CLARITY images of different conditions (Challenged to Control).
In the final experiment, CLARITY images were registered to other CLARITY images of the same condition (Challenged to Challenged).
Registration results from the Image-LDDMM and Mask-LDDMM pipelines were compared to a MI-BSpline pipeline described in an upcomming work by Tomer \textit{et al.}
The pipeline registers images using their Mutual Information under a B-Spline spatial transformation.
The results of these experiments are summarized in Fig.~\ref{fig:expriments}-\ref{fig:summary}.

\subsection{CLARITY-ARA}
Fig.~\ref{fig:clarityToAra} and the green lines in Fig.~\ref{fig:summary} show the results of CLARITY to ARA registrations.
The Mask-LDDMM pipeline consistently outperformed the MI-BSpline and Image-LDDMM pipelines in surface and landmark distances. 
Interestingly, the Image-LDDMM pipeline consistently yielded higher MI values than the Mask-LDDMM and MI-BSpline pipelines.
Despite this, visual inspection of Image-LDDMM results in Fig.~\ref{fig:clarityToAra} revealed that the alignments were relatively poor.
This can likely be attributed to the great difference in appearance between the Nissl-stained ARA and CLARITY images.

\subsection{CLARITY-CLARITY between conditions}
Fig.~\ref{fig:betweenConditions} and the blue lines in Fig.~\ref{fig:summary} show CLARITY to CLARITY registration results between conditions.
The MI-BSpline method did better than the Image and Mask-LDDMM pipelines in MI and median landmark error.
Mask-LDDMM, Image-LDDMM, and MI-BSpline gave similar surface distance results.

\subsection{CLARITY-CLARITY within a condition}
\begin{figure}[th]
 \includegraphics[width=\textwidth]{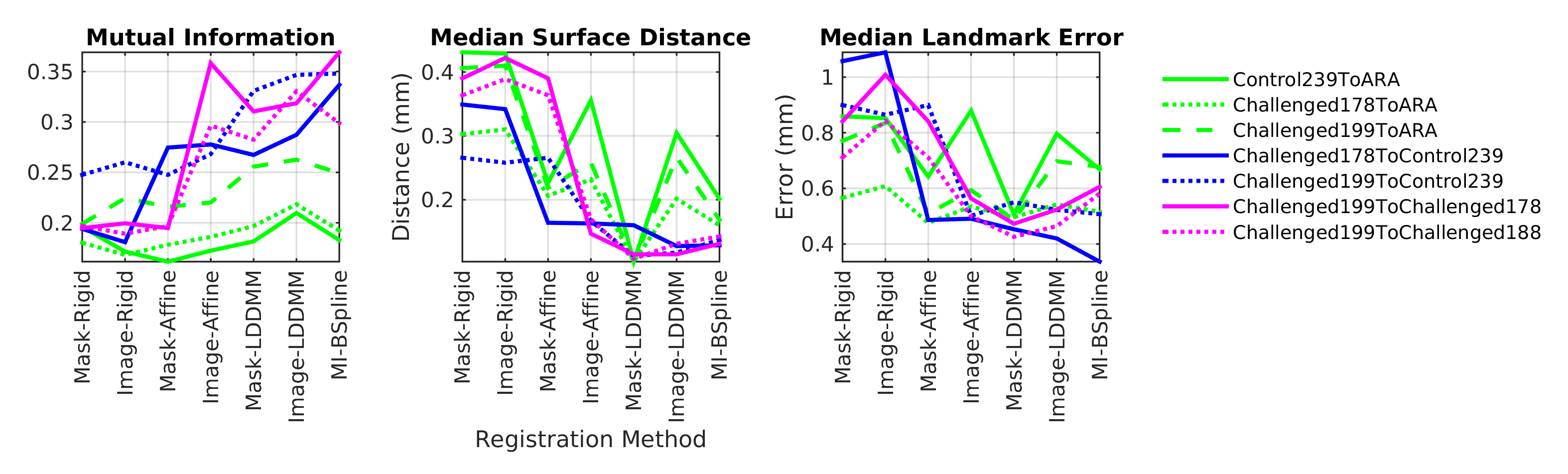}
 \caption{ A summary of metric values for all experiments over each registration method. }
 \label{fig:summary}
\end{figure}

Fig.~\ref{fig:withinConditions} and the magenta lines in Fig.~\ref{fig:summary} show the registration results of CLARITY to CLARITY registrations within a condition.
As in the inter-condition registrations, Image-LDDMM yielded better MI values than Mask-LDDMM. 
But unlike the inter-condition registrations, Image-LDDMM and Mask-LDDMM had lower median landmark errors than the MI-BSpline method.
Once again Mask-LDDMM, Image-LDDMM, and MI-BSpline gave comparable surface distance results.

\section{CONCLUSION}
In most cases, the Mask-LDDMM pipeline outperformed both Image-LDDMM and MI-BSpline in aligning CLARITY brains with the ARA.
The MI-BSpline pipeline gave better results than Image-LDDMM in CLARITY-CLARITY transforms between conditions.
Image-LDDMM outperformed MI-BSpline in CLARITY-CLARITY transforms within a condition.
However, there are some limitations to these findings.
Human error in landmark placement may have been a factor in these results.
Furthermore, the pipelines were tested on brains without missing data.  
When only partial data is available, complete brain masks cannot be constructed and Mask-LDDMM should not be used.

\acknowledgments 
This work was graciously supported by the Defense Advanced Research Projects Agency (DARPA) SIMPLEX program through SPAWAR contract N66001-15-C-4041 and DARPA GRAPHS N66001-14-1-4028. 


\bibliography{kwame} 
\bibliographystyle{spiebib} 

\end{document}